\documentclass[twocolumn,superscriptaddress,10pt]{revtex4-1}
\usepackage[utf8]{inputenc}
\usepackage{amsmath}
\usepackage{amssymb}
\usepackage{graphicx}
\usepackage{subfigure}
\usepackage{dcolumn}
\usepackage{enumerate}
\usepackage{textcomp}
\newcommand{\QE}{\mbox{Quantum-Espresso}}
\newcommand{\Elk}{\mbox{Elk}}
\newcommand{\cminv}{$\textrm{cm}^{-1}$}
\newcommand{\mkm}{\textmu{}m}
\newcommand{\mks}{\textmu{}s}
\newcommand{\Biz}{$\textrm{Bi}^{0}$}
\newcommand{\Bipi}{$\textrm{Bi}^{+}$}
\newcommand{\Bipii}{$\textrm{Bi}^{2+}$}
\newcommand{\Bipiii}{$\textrm{Bi}^{3+}$}
\newcommand{\Biiiz}{$\textrm{Bi}_2^{0}$}
\newcommand{\Biiim}{$\textrm{Bi}_2^{-}$}
\newcommand{\BiO}{$\textrm{BiO}$}
\newcommand{\BiVSi}%
{\mbox{$\textrm{Bi}\cdots\equiv\!\textrm{Si}\!\relbar\!\textrm{Si}\!\equiv$}}
\newcommand{\BiVGe}%
{\mbox{$\textrm{Bi}\cdots\equiv\!\textrm{Ge}\!\relbar\!\textrm{Ge}\!\equiv$}}
\newcommand{\SiODCii}{\mbox{$=\!\textrm{Si}$}}
\newcommand{\SiSi}{\mbox{$\equiv\!\textrm{Si}\!\relbar\!\textrm{Si}\!\equiv$}}
\newcommand{\GeGe}{\mbox{$\equiv\!\textrm{Ge}\!\relbar\!\textrm{Ge}\!\equiv$}}
\newcommand{\SiOii}{\mbox{$\textrm{SiO}_2$}}
\newcommand{\GeOii}{\mbox{$\textrm{GeO}_2$}}
\newcommand{\SiOiiBi}{\mbox{$\textrm{SiO}_2\textrm{:Bi}$}}
\newcommand{\GeOiiBi}{\mbox{$\textrm{GeO}_2\textrm{:Bi}$}}
\newcommand{\Term}[4]{\mbox{${}^{#1}{\textrm{#2}}_{#3}{#4}$}}

\newlength{\Li}
\begin{document}
\title{%
Centers of near-IR luminescence 
in Bi-doped \mbox{SiO$_2$}{} and \mbox{GeO$_2$}: \\
First-principle modeling and experimental data analysis
}
\author{V.~O.~Sokolov}\email{Corresponding author: vence.s@gmail.com}
\affiliation{Fiber ics Research Center of the Russian Academy of
Sciences \\ 38 Vavilov Street, Moscow, 119333, Russia
}
\author{V.~G.~Plotnichenko}
\affiliation{Fiber Optics Research Center of the Russian Academy of
Sciences \\ 38 Vavilov Street, Moscow, 119333, Russia
}
\affiliation{Moscow Institute of Physics and Technology \\
9 Institutskii per., Dolgoprudny, Moscow Region, 141700, Russia
}
\author{E.~M.~Dianov}
\affiliation{Fiber Optics Research Center of the Russian Academy of
Sciences \\ 38 Vavilov Street, Moscow, 119333, Russia
}
\begin{abstract}
First-principle study of possible bismuth-related centers in \SiOii{} and
\GeOii{} hosts is performed and the results are compared with the experimental
data. The following centers are modeled: trivalent and divalent Bi
substitutional centers; \BiO{} interstitial molecule; interstitial ion, \Bipi,
and atom, \Biz; \BiVSi{} and \BiVGe{} complexes formed by interstitial Bi atoms
and glass intrinsic defects, \SiSi{} or \GeGe{} oxygen vacancies; interstitial
dimers, \Biiiz{} and \Biiim. Experimental data available on bismuth-related IR
luminescence in \SiOiiBi{} and \GeOiiBi{} glasses, visible (red) luminescence in
\SiOiiBi{} glass and luminescence excitation are analyzed. A comparison of
calculated spectral properties of bismuth-related centers with the experimental
data shows that the IR luminescence in \SiOiiBi{} and \GeOiiBi{} is most likely
caused by \BiVSi{} and \BiVGe{} complexes, and divalent Bi substitutional center
is probably responsible for the red luminescence in \SiOiiBi.
\end{abstract}

\maketitle

\section{%
Introduction
}
\label{sec:Introduction}
Bismuth-doped glasses and optical fibers based on such glasses attract a
considerable interest caused by characteristic broadband IR luminescence in the
range from 1.0 up to 1.7~\mkm{} and even beyond, covering almost entire
telecommunications window. Bismuth-doped glasses become a promising active
medium. Such glasses and fibers are studied intensively and used successfully in
fiber amplification and lasing (see e.g. the reviews \cite{Dianov09, Dianov13}).
The origin of IR luminescence centers is still not clear. However for the last
few years an opinion has received a recognition that the luminescence is very
likely caused by subvalent Bi centers \cite{Peng11} (see \cite{Dianov10} as
well). In particular, monovalent Bi centers are of special interest.

In understanding the origin of the luminescence centers simple hosts are of a
specific interest since those allow one to get more unambiguous data in both
experimental and theoretical studies. For example, such simple crystals as
halides of monovalent metals, are convenient \cite{Plotnichenko13, Sokolov13}.
As for fiber optics applications, in this regard the results of investigation of
the luminescence in optical fibers with bismuth-doped silica (\SiOii) and
germania (\GeOii) glass core not containing any other dopants \cite{Bufetov10,
Bufetov11a, Bufetov11b, Bufetov12a, Bufetov12b, Bufetov13} are undoubtedly of a
prime importance. The main results concerning the spectral properties of
\SiOiiBi{} are confirmed later independently (see e.g. Refs.~\cite{Trukhin13a,
Trukhin13b}).  For further analysis the following spectral properties of
bismuth-related centers in \SiOiiBi{} and \GeOiiBi{} discovered in
Refs.~\cite{Bufetov10, Bufetov11a, Bufetov11b, Bufetov12a, Bufetov12b,
Bufetov13, Trukhin13a, Trukhin13b} seem to be of a principal interest:
\begin{enumerate}[i.]
\item \label{itm:i} luminescence around 1.43~\mkm{} is excited in
\SiOiiBi{} by absorption near 1.43, 0.83, 0.42, 0.37 and 0.24~\mkm{}
\cite{Bufetov11a, Bufetov11b, Bufetov12a, Trukhin13b};
\item \label{itm:ii} luminescence around 1.67~\mkm{} is excited in
\GeOiiBi{} due to absorption near 1.65, 0.93, 0.46 and 0.40~\mkm{}
\cite{Bufetov11b, Bufetov12a, Bufetov13};
\item \label{itm:iii} luminescence near 0.83~\mkm{} is excited in \SiOiiBi{}
by absorption near 0.83 and 0.42~\mkm{} \cite{Bufetov11b};
\item \label{itm:iv} luminescence near 0.95~\mkm{} is excited in \GeOiiBi{}
by absorption near 0.93 and 0.46~\mkm{} \cite{Bufetov11b};
\item \label{itm:v} luminescence in the red range, 0.60--0.65~\mkm, is
excited in \SiOiiBi{} due to absorption near 0.48, 0.37 and $\lesssim
0.29$~\mkm{} \cite{Bufetov11b, Trukhin13b}; no such luminescence is observed in
\GeOiiBi{} \cite{Bufetov11b};
\item \label{itm:vi} an electron-hole recombination excitation mechanism
contributes to the red luminescence, but not to the IR one in \SiOiiBi{}
under intensive UV illumination \cite{Trukhin13b};
\item \label{itm:vii} red luminescence in \SiOiiBi{} is thermally quenched
at temperatures $\gtrsim 450$~K, while the IR luminescence is not quenched up to
the temperature of 700~K \cite{Bufetov12a, Trukhin13b};
\item \label{itm:viii} the lifetimes of the states responsible for the
luminescence near 1.43 and 0.83~\mkm{} in \SiOiiBi{} and near 1.67~\mkm{} in 
\GeOiiBi, are 640, 40 and 500~\mks, respectively \cite{Bufetov11a, Bufetov13}
(no data available on lifetime of the state responsible for 0.95~\mkm{}
luminescence in \GeOiiBi).
\end{enumerate}

Qualitative similarity of the absorption and luminescence spectra in \SiOiiBi{}
and in \GeOiiBi{} should be particularly emphasized. In Refs.~\cite{Bufetov10,
Bufetov11a, Bufetov11b, Bufetov12a, Bufetov12b, Bufetov13}, where this fact was
revealed, the suggestion was made that the centers responsible for the IR
luminescence in bismuth-doped \SiOii{} and \GeOii{} glasses have a common origin
and a similar structure, and the levels and transitions schemes of these centers
were proposed.

In the present work we report the results of computer modeling of possible
bismuth-related centers in bismuth-doped \SiOii{} and \GeOii{} glasses and try
to interpret the above-listed experimental data.

\section{%
Modeling of bismuth-related centers
}
\label{sec:Modeling}
The modeling of bismuth-related centers in silica and germania glass network was
performed using periodical network models. $2 \times 2 \times 2$ and $3 \times 3
\times 2$ supercells of \SiOii{} and \GeOii{} lattices of $\alpha$ quartz
polymorph structure were chosen as initial models of perfect networks. The
supercells contained, respectively, 24 or 54 \SiOii{} (\GeOii) groups (72 or 162
atoms). Using ab~initio Car-Parrinello molecular dynamics \cite{Car85} the
system formed by supercells was heated to temperature as high as 2000~K (\SiOii)
or 1500~K (\GeOii), maintained at this temperature until the equilibrium atom
velocities distribution was reached and then cooled to 300~K. Periodical models
of \SiOii{} and \GeOii{} networks based on final supercell configurations were
applied to study the bismuth-related centers. Bi atoms (e.g. an interstitial
atom) were placed in the central region of the supercell. As well an oxygen
vacancy, \SiSi{} or \GeGe, was formed there by a removal of one of the O atoms.
Charged centers were simulated changing the total number of electrons in the
supercell. Equilibrium configurations of Bi centers were found by a subsequent
Car-Parrinello MD and complete optimization of the supercell parameters and
atomic positions by the gradient method.

All calculations of \SiOii{} and \GeOii{} network models and configuration of
bismuth-related centers were performed using \QE{} package \cite{QE} in the
plane wave basis in generalized gradient approximation of density functional
theory with ultra-soft PAW \cite{PAW} pseudopotentials. The pseudopotential
sources were taken from pslibrary~v.~0.3.0 pseudopotential library
\cite{pslibrary}. PBE density functional \cite{PBE} was used both in building
the pseudopotentials and in calculations. To test the approach, \SiOii{} and
\GeOii{} lattice parameters were calculated for $\alpha$ quartz unit cell and
for supercells with both atomic positions and cell parameters completely
optimized. The convergence was tested with respect to the plane wave cutoff
energy and the $k$ points grid choice. The energy cutoff $\gtrsim 950$~eV and
the number of $k$ points $\geq 64$ in the irreducible part of unit cell
Brillouin zone were found to be enough to converge the total energy within
$10^{-3}$~eV per atom and to reproduce the experimental lattice parameters with
a relative accuracy of $\lesssim 5 \times 10^{-3}$. The geometry of supercells
was reproduced with a relative accuracy better than $2 \times 10^{-2}$ with only
$\Gamma$ point of the supercell taken into account and better than $1 \times
10^{-3}$ using 8 $k$ points in the supercell in irreducible part of the
supercell Brillouin zone. The total energy convergence was not worse than that
in case of the unit cell.

Configurations of bismuth-related centers obtained by this means were used to
calculate the electron localization functions using the programs from \QE{}
package, to calculate and analyze the electron density distribution and
effective charges of atoms by Bader's method(bader~v.\,0.28 code \cite{bader}),
and to calculate the absorption spectra of the centers by Bethe-Salpeter
equation method based on all-electron full-potential linearized augmented-plane
wave approach \cite{LAPW}. The latter calculations were performed with \Elk{}
code \cite{Elk} in the local spin density approximation with PW-CA functional
\cite{LSDA-PW, LSDA-CA}. Spin-orbit interaction essential for Bi-containing
systems was taken into account. Scissor correction was used to calculate
transition energies. The scissor value was found basing on calculations with
modified Becke-Johnson exchange-correlation potential known to yield accurate
band gaps in wide-band-gap insulators, sp semiconductors, and transition metal
oxides \cite{Becke06, Tran09, Tran11}. The non-overlapping muffin-tin (MT)
spheres of maximal possible radii $R^\textrm{MT}$ were used. Convergence of the
results was tested with respect to plane-wave cutoff energy, the angular
momentum cutoff for the MT density and potential, and the $k$ points grid
choice. The plane-wave cutoff, $k_{max}$, was determined by $R^\textrm{MT}_{min}
\cdot k_{max} = 7$ relation with $R^\textrm{MT}_{min}$ being the smallest MT
radius. The angular momentum cut-off was taken to be $l = 10$. The
self-consistent calculations were performed on $4\times 4\times 4$ grid of $k$
points uniformly distributed in irreducible part of the supercell Brillouin
zone. Further increasing the cutoff and $k$ points density
did not change the results significantly. The total energy self-consistence
tolerance was taken to be $10^{-3}$~eV per atom. More dense $k$ points grid,
$8\times 8\times 8$, was used to calculate dipole matrix elements for optical
spectra.

Configurational coordinate curves of bismuth-related centers were calculated
in a simple model restricted to the lowest excited states basing on absorption
spectra dependence on Bi atom(s) displacement along three mutual orthogonal
directions. In spite of the fact that the model is inherently approximate, it
shows that in all Bi interstitial-related centers under consideration the Stokes
shift corresponding to a transition between the first excited state and the
ground one do not exceed the accuracy of the excited state energy calculation.
Hence it is reasonable enough to estimate the IR luminescence wavelengths in
such centers by taking the Stokes shift to be zero. On the contrary, in all 
centers with oxygen-bonded Bi atom(s) the corresponding Stokes shift turns out
to be large. So in such centers we can estimate only roughly the IR luminescence
wavelengths by analogy with known centers in other hosts.

In our approach only approximate calculations of absorption intensities or
excited states lifetimes are possible. Much higher density of the $k$ points
grid and, in general, a considerably larger supercell are required to achieve an
accuracy comparable to the experimental data. Nevertheless, a comparison of the
results of test calculations performed in our approach and the experimental data
available for exciton absorption in \SiOii{} and for \SiODCii{} oxygen-deficient
center absorption in silica glass showed that the relative absorption intensity
is reproduced with an accuracy not worse than an order of magnitude. In view of
the above remark about the excited states one might expect that the relative
lifetimes of the states responsible for the luminescence are estimated with
nearly the same accuracy.

\section{%
Results of calculations on bismuth-related centers in SiO$_2$ and GeO$_2$ hosts
}
\label{sec:Calculation}
Basing on the assumptions concerning subvalent Bi states we studied several
bismuth-related centers in \SiOiiBi{} and \GeOiiBi{} networks, as follows:
\begin{itemize}
\item[---] trivalent Bi substitutional center;
\item[---] divalent Bi substitutional center;
\item[---] \BiO{} interstitial molecule;
\item[---] interstitial single-changed ion, \Bipi, and interstitial atom, \Biz;
\item[---] \BiVSi{} and \BiVGe{} complexes formed by interstitial Bi atom and
glass intrinsic defect, \SiSi{} or \GeGe{} oxygen vacancy;
\item[---] interstitial Bi dimers, \Biiiz{} and \Biiim.
\end{itemize}
These centers correspond to different bismuth oxidation levels (valence states),
namely, $\textrm{Bi}^\textrm{III}$ (trivalent Bi substitutional center),
$\textrm{Bi}^\textrm{II}$ (divalent Bi substitutional center and \BiO{}
molecule), $\textrm{Bi}^\textrm{I}$ (interstitial single-changed ion, \Bipi,
and \BiVSi{} or \BiVGe{} complexes), and to completely reduced Bi (interstitial
atom, \Biz, and \Biiiz{} and \Biiim{} dimers). We believe that our choice covers
sufficiently wide range of the most probable bismuth-related centers in
\SiOiiBi{} and \GeOiiBi{} glasses.

\subsection{%
Trivalent Bi substitutional center
}
\label{sec:Bi^3+}
According to our modeling of Bi substitutional centers in \SiOii{} and \GeOii{}
networks, Bi atoms occupying, respectively, Si or Ge sites, can form
substitutional centers of two types, threefold coordinated Bi atoms bonded by
three bridging O atoms with Si (Ge) atoms, and fourfold coordinated Bi atoms in
bi-pyramidal configuration bonded by four bridging O atoms with Si (Ge) atoms.
In addition to bridging O atoms, in both cases there are also O atoms completing
the Bi atoms coordination to fivefold or sixfold ones. These O atoms interact
slightly with Bi atoms forming only very weak bonds. Threefold (to be more
specific, $3 + 3$- or $3 + 2$-fold) coordination of atoms is typical of
trivalent Bi compounds and of impurity bismuth centers in many hosts (see e.g.
\cite{Blasse68, Blasse84}). Fourfold ($4 + 2$- or $4 + 1$-fold) coordination of
Bi atoms is known to occur in some crystalline bismuth compounds (see e.g.
\cite{Rossell92, Dinnebier02}). Formation of additional bonds of Bi atom in such
centers becomes possible owing to threefold coordinated O atom occurring
anywhere in the network. Four-coordinated Bi atoms turn out to be rather
unstable: in the modeling they are either easily transformed into threefold
coordinated atoms or even not formed at all.
\begin{figure}
\begin{center}
\includegraphics[width=8.75cm, bb=-10 -10 1110 1110]{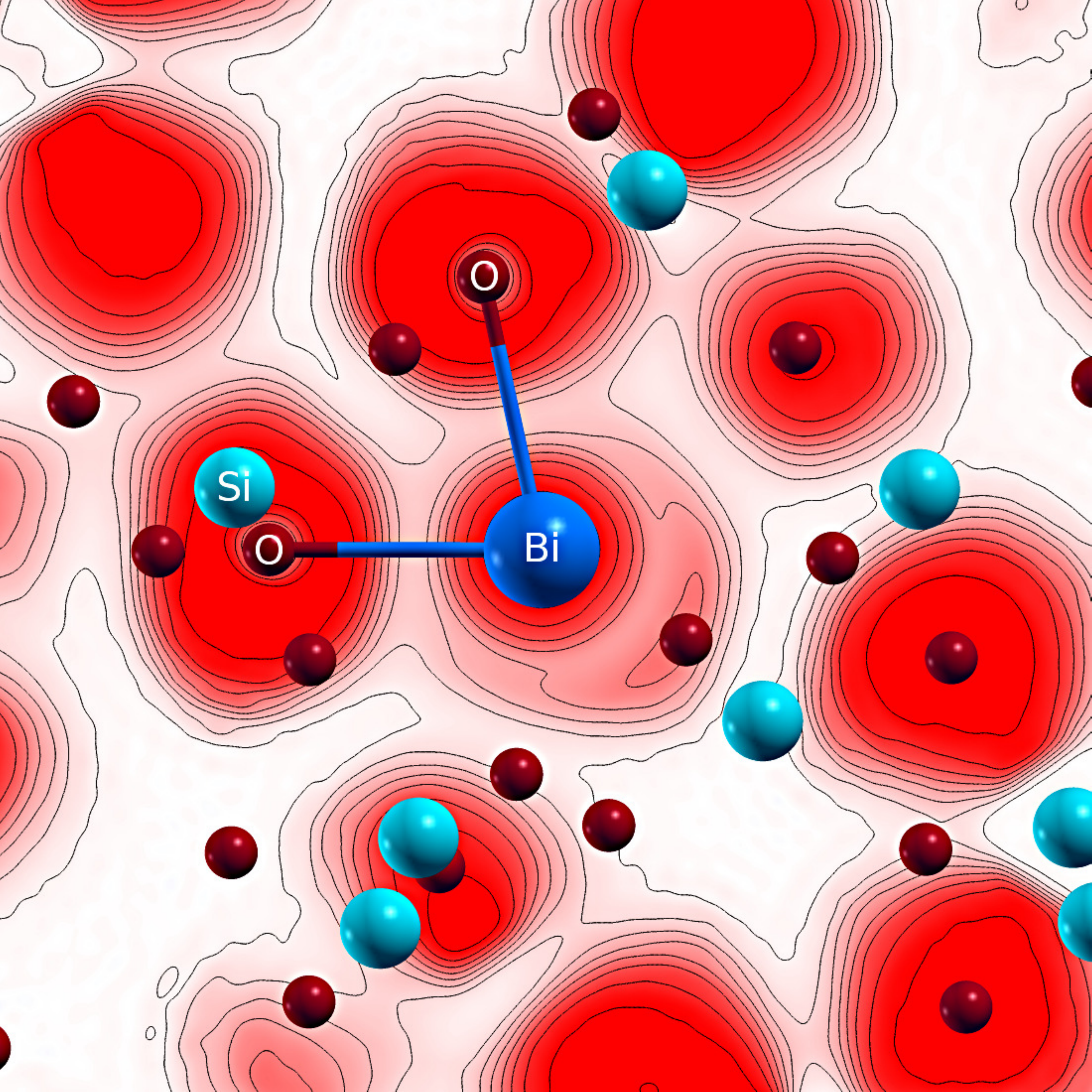}
\end{center}
\caption{%
Calculated electron density map of \Bipii{} center in \SiOii. The map plane goes
through the Bi atom and two bridging O atoms.
}
\label{fig:Bi+2_SiO2_ELF}
\end{figure}

In the centers with threefold coordinated Bi atoms the
\mbox{$\textrm{Bi}\!\relbar\!\textrm{O}$} distances are found to be
approximately 0.213~nm for three bridging O atoms and 0.34 -- 0.36~nm for
complementary O atoms. The angles between Bi$\relbar$O bonds are about
$94.6^\circ$ for bridging O atoms. In the centers with fourfold coordinated Bi
atoms the \mbox{$\textrm{Bi}\!\relbar\!\textrm{O}$} distances are
approximately 0.209 and 0.230~nm for equatorial and axial bridging O atoms,
respectively, and about 0.33~nm for complementary O atoms. The
\mbox{$\textrm{O}\!\relbar\!\textrm{Bi}\!\relbar\!\textrm{O}$} valence angles
are about $92.2^\circ$ and $160.8^\circ$ for equatorial and axial bridging O
atoms, respectively.
\begin{figure*}
\subfigure[]{%
\includegraphics[width=3.40cm, bb=100 -15 290 815]{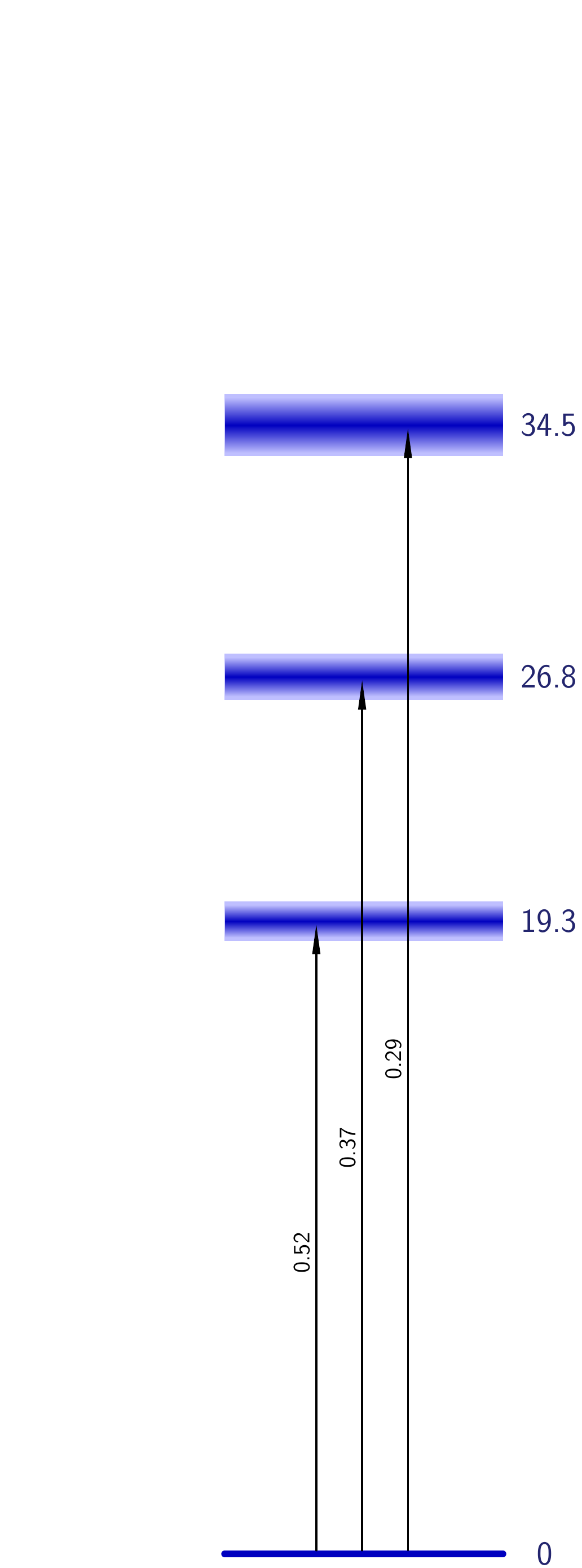}
\label{fig:Bi+2_SiO2_levels}
}
\subfigure[]{%
\includegraphics[width=3.40cm, bb=100 -15 290 815]{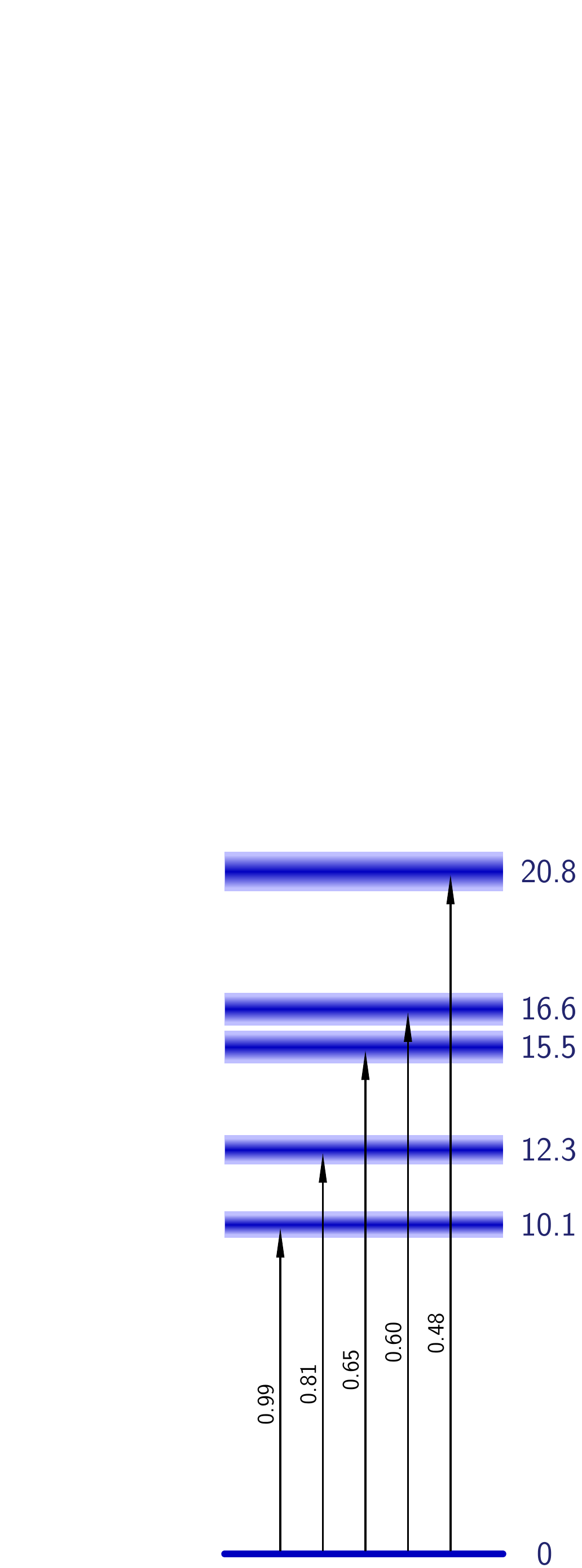}
\label{fig:Bi+_SiO2_levels}
}
\subfigure[]{%
\includegraphics[width=3.40cm, bb=100 -15 290 815]{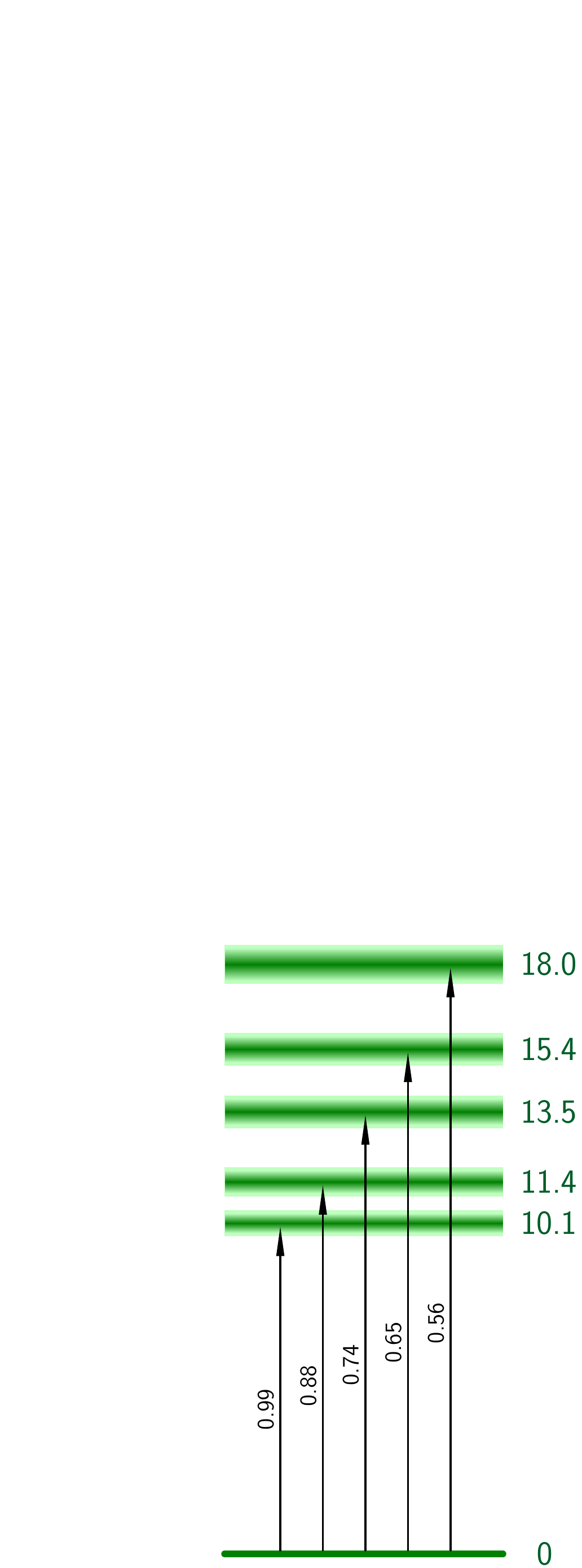}
\label{fig:Bi+_GeO2_levels}
}
\subfigure[]{%
\includegraphics[width=3.40cm, bb=100 -15 290 815]{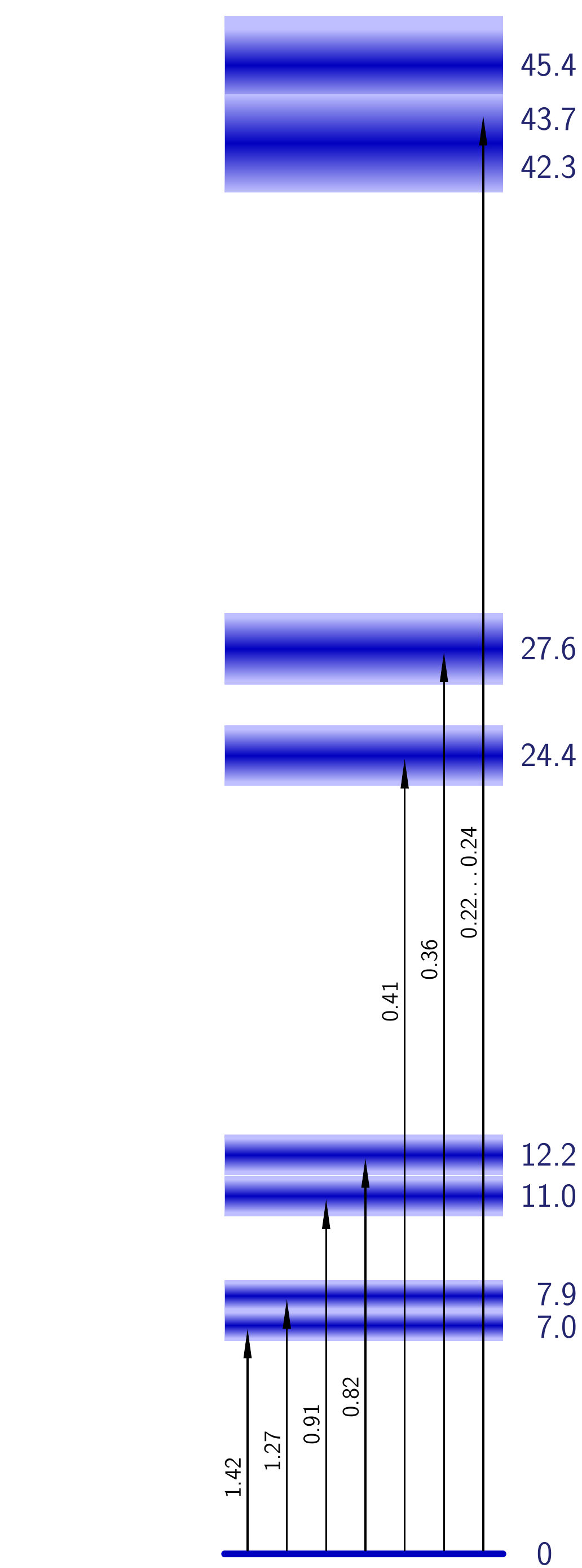}
\label{fig:Bi0_SiSi_levels}
}
\subfigure[]{%
\includegraphics[width=3.40cm, bb=100 -15 290 815]{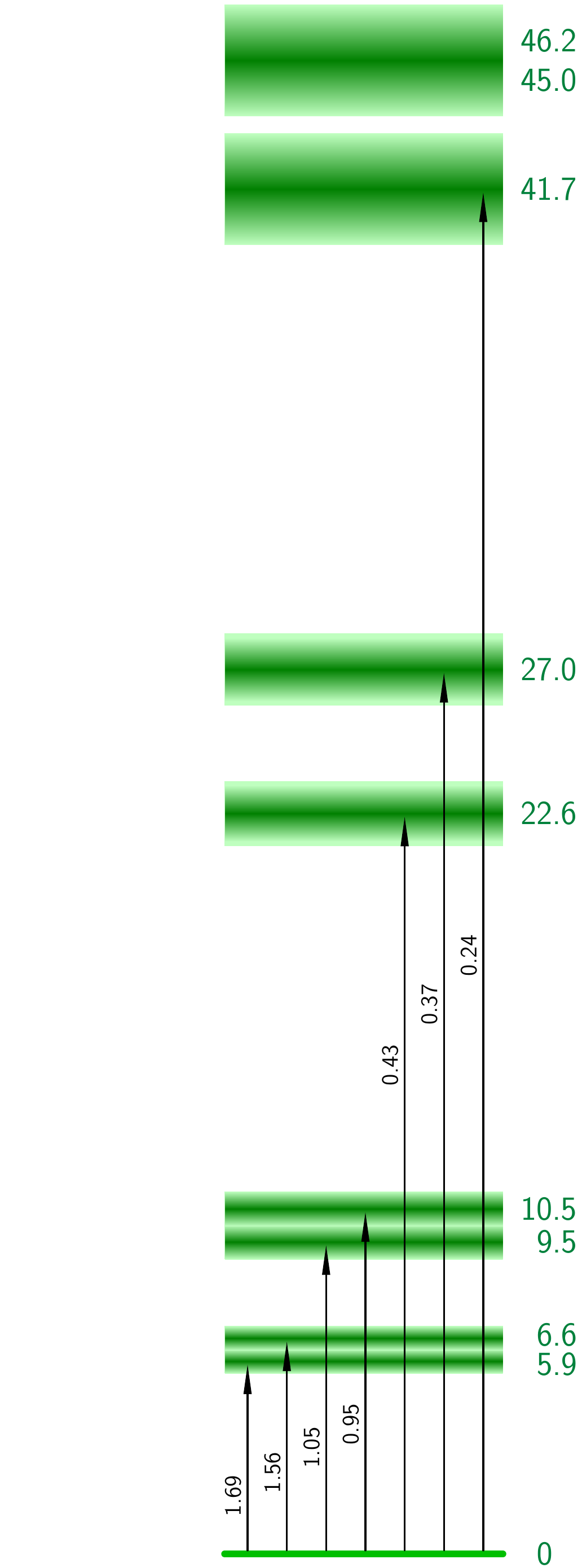}
\label{fig:Bi0_GeGe_levels}
}
\caption{%
Calculated levels and transitions schemes of bismuth-related centers:
\subref{fig:Bi+2_SiO2_levels}~\Bipii{} center in \SiOii,
\subref{fig:Bi+_SiO2_levels}~interstitial \Bipi{} center in \SiOii,
\subref{fig:Bi+_GeO2_levels}~interstitial \Bipi{} center in \GeOii,
\subref{fig:Bi0_SiSi_levels}~\BiVSi{} center in \SiOii,
\subref{fig:Bi0_GeGe_levels}~\BiVGe{} center in \GeOii.
Level energies are given in $10^3$~\cminv, transition wavelengths in \mkm.
}
\label{fig:Bi_centers_levels}
\end{figure*}
Spectral properties of trivalent substitutional Bi centers (usually named as
\Bipiii{} impurity centers) are much studied (see e.g. \cite{Blasse68,
Blasse84}). Absorption and luminescence bands corresponding to
\Term{1}{S}{0}{}$\,\longleftrightarrow\,$\Term{3}{P}{1}{}{} transition in
\Bipiii{} ion are commonly observed \cite{Moore58, George85, Wahlgren01}.
Sometimes the absorption corresponding to the
\Term{1}{S}{0}{}$\,\longrightarrow\,$\Term{1}{P}{1}{} transition is present as
well. Wavelengths of absorption and luminescence bands are subject to wide
variations depending on the host: luminescence in the 0.28--0.55~\mkm{} range is
excited by absorption in the 0.23--0.33~\mkm{} range. Lifetimes of excited
states responsible for the luminescence are typically of 1--5~\mks{} in
different hosts. The results of our modeling suggest that these variations may
be explained by a prevailing coordination of trivalent Bi. E.g. in \SiOii{} the
most intensive absorption band of the threefold coordinated Bi atom is found to
be near 0.26~\mkm{} but the same band of the fourfold coordinated Bi atom turns
out to be shifted to 0.22~\mkm. In \GeOii{} the absorption bands of threefold
and fourfold coordinated Bi atoms occur near 0.24 and 0.23~\mkm, respectively.
Stokes shift turn out to be large in all these cases. Hence we can do no more
than suggest, by analogy with the experimental data available \cite{Blasse68,
Blasse84}, that the luminescence in the 0.3--0.4~\mkm{} range may be excited in
the above-mentioned absorption bands. 

\subsection{%
Divalent Bi substitutional center
}
\label{sec:Bi^2+}
The modeling revealed that in \SiOii{} network the divalent bismuth forms
twofold coordinated Bi atoms bonded by bridging O atoms with Si atoms. In
\GeOii{} such twofold coordinated Bi atom does not occur since \GeOii{} network
readily transforms resulting in a formation of threefold coordinated Bi atom
(the above-described trivalent substitutional center), sixfold coordinated Ge
atom, and threefold coordinated O atom.

Calculation of electron density distribution by Bader's method proves the
effective charge of twofold coordinated Bi atom to be
$+1.316\left|\textrm{e}\right|$. The effective charges of bridging O atoms
bonding Bi atom with Si ones turn out to be $-1.629\left|\textrm{e}\right|$, and
those of Si atoms are $+3.927\left|\textrm{e}\right|$. In perfect \SiOii{}
network the effective charges of O and Si atoms calculated using the same
approach are found to be $-1.965\left|\textrm{e}\right|$ and
$+3.930\left|\textrm{e}\right|$, respectively. Hence an extra charge as large as
$\approx +1.99\left|\textrm{e}\right|$ turns out to be localized in Bi atom and
bridging O atoms, suggesting that Bi atom is divalent in this center. In
Fig.~\ref{fig:Bi+2_SiO2_ELF} the distribution of electron density around twofold
coordinated Bi atom is shown. One can recognize two
\mbox{$\textrm{Bi}\!\relbar\!\textrm{O}$} bonds formed.

Calculated energy levels and transitions in \Bipii{} substitutional center
(Fig.~\ref{fig:Bi+2_SiO2_levels}) agree well both with the data available on
absorption and luminescence of divalent Bi centers \cite{Blasse94, Blasse97,
Srivastava98} and with the spectra measured in fibers and fiber preforms with
\SiOii{} core \cite{Bufetov11b, Trukhin13b}. The ground state of free \Bipii{}
ion is known to be \Term{2}{P}{1/2}{} and the first excited state with the
energy about 20800~\cminv{} is \Term{2}{P}{3/2}{}{} \cite{Moore58, Wahlgren01,
Dolk02}. In crystal field this excited state is split in two sub-levels,
\Term{2}{P}{3/2}{\left(1\right)}{} and \Term{2}{P}{3/2}{\left(2\right)}.
According to Refs.~\cite{Blasse94, Blasse97, Srivastava98}, the absorption
bands characteristic of \Bipii{} correspond to
\Term{2}{P}{1/2}{}$\,\longrightarrow\,$\Term{2}{P}{3/2}{\left(1\right)},
\Term{2}{P}{1/2}{}$\,\longrightarrow\,$\Term{2}{P}{3/2}{\left(2\right)}, and
\Term{2}{P}{1/2}{}$\,\longrightarrow\,$\Term{2}{S}{1/2}{} transitions.
Luminescence in 0.55--0.65~\mkm{} range (in different hosts) excited in these
absorption bands corresponds to
\Term{2}{P}{3/2}{\left(1\right)}$\,\longrightarrow\,$\Term{2}{P}{1/2}{}
transition. In our calculation the absorption bands corresponding to three
above-mentioned transitions are near 0.52, 0.37, and $< 300$~\mkm, respectively
(Fig.~\ref{fig:Bi+2_SiO2_levels}). As in the case of trivalent Bi centers,
Stokes shift turns out to be large in divalent Bi substitutional center, and
so we can estimate only the luminescence wavelength as 0.6--0.7~\mkm, by analogy
with the data of Refs.~\cite{Blasse94, Blasse97, Srivastava98}.

\subsection{%
Interstitial \BiO{} molecule
}
\label{sec:BiO_molecule}
We have found in our modeling that in \SiOii{} network there is an equilibrium
position of \BiO{} molecule in the interstitial site formed by six-member rings
of SiO$_4$ tetrahedra. In such a position the \BiO{} molecule is placed between
two neighboring rings being aligned along the interstitial axis. This position
of \BiO{} molecule is found to be quite stable: being forcedly declined from the
interstitial axis or shifted aside, the molecule do not enter into reaction with
the surrounding atoms and returns to the initial position. The present modeling
confirms the results of our previous calculation  (see Fig.~2 in
Ref.~\cite{Sokolov11}) performed in a cluster model of \SiOii{} network using
quantum-chemical methods. This is very distinct from the case of aluminosilicate
glasses \cite{Sokolov09}, where the interstitial position of \BiO{} molecule
turns out to be completely unstable, and \BiO{} molecule reacts readily with
neighboring atoms forming threefold coordinated Bi atom.
\begin{figure}
\includegraphics[width=8.75cm, bb=5 0 485 490]{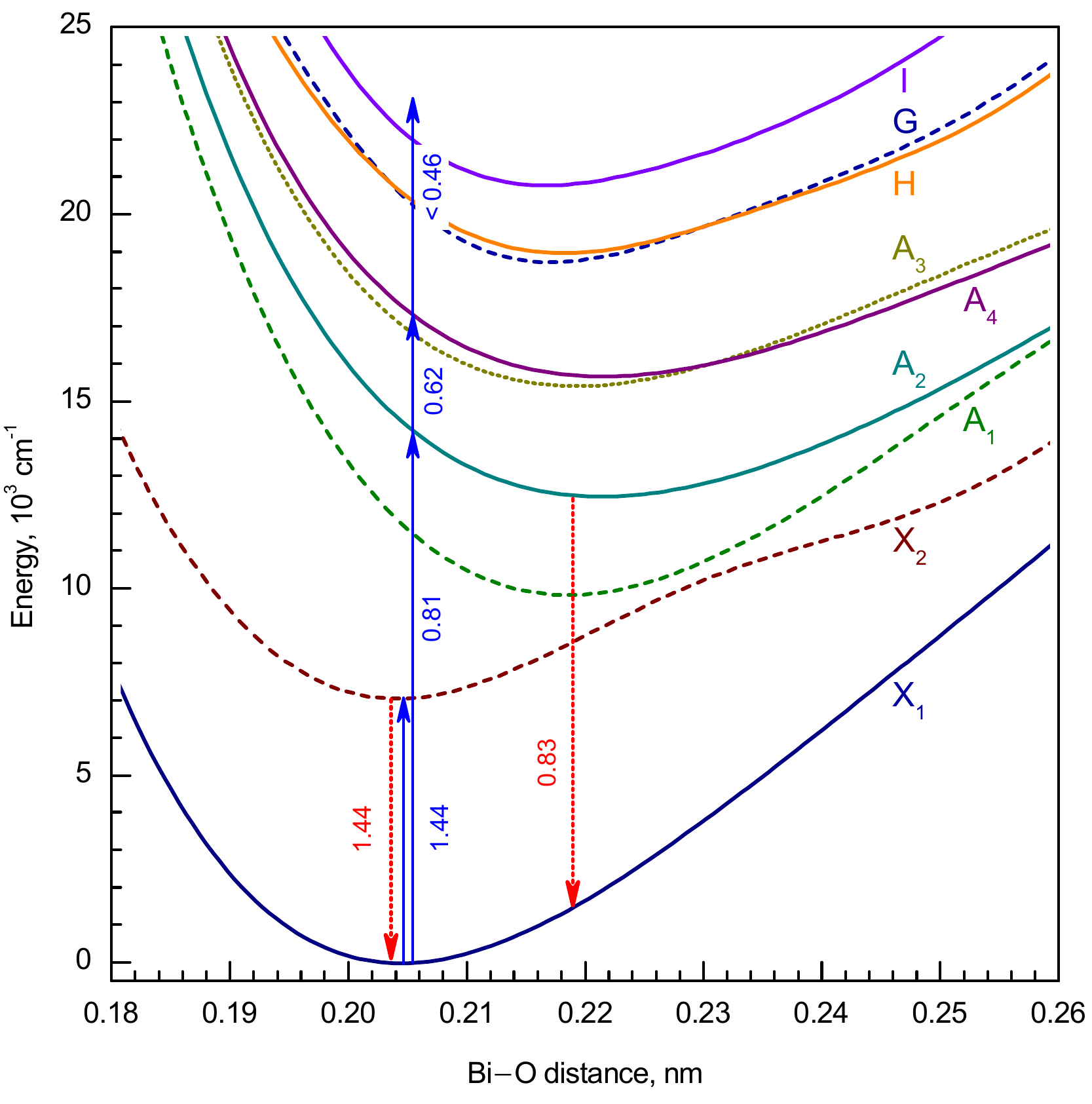}
\caption{%
Total energy curves and transitions in \BiO{} molecule according to data of
Refs.~\cite{Alekseyev94, Shestakov98, DiatomicMolecules}.
}
\label{fig:BiO_levels}
\end{figure}

In contrast to \SiOii, interstitial \BiO{} molecule is not found to occur in
\GeOii{} network in our modeling. Quite as it is in aluminosilicate host, the
molecule reacts with \GeOii{} network with threefold coordinated Bi atom formed.

Spectral properties of \BiO{} molecule are well understood \cite{Shestakov98,
Alekseyev94, DiatomicMolecules}. Total energy curves of \BiO{} molecule obtained
from the configuration interaction calculations with spin-orbit interaction
taken into account \cite{Alekseyev94} and corrected to achieve more close
agreement with the experimental data \cite{Shestakov98, DiatomicMolecules} are
given together with possible transitions in Fig.~\ref{fig:BiO_levels} (see
as well Fig.~1 in Ref.~\cite{Sokolov11}).
\begin{figure*}
\subfigure[]{%
\includegraphics[width=8.75cm, bb=-10 -10 1110 1110]{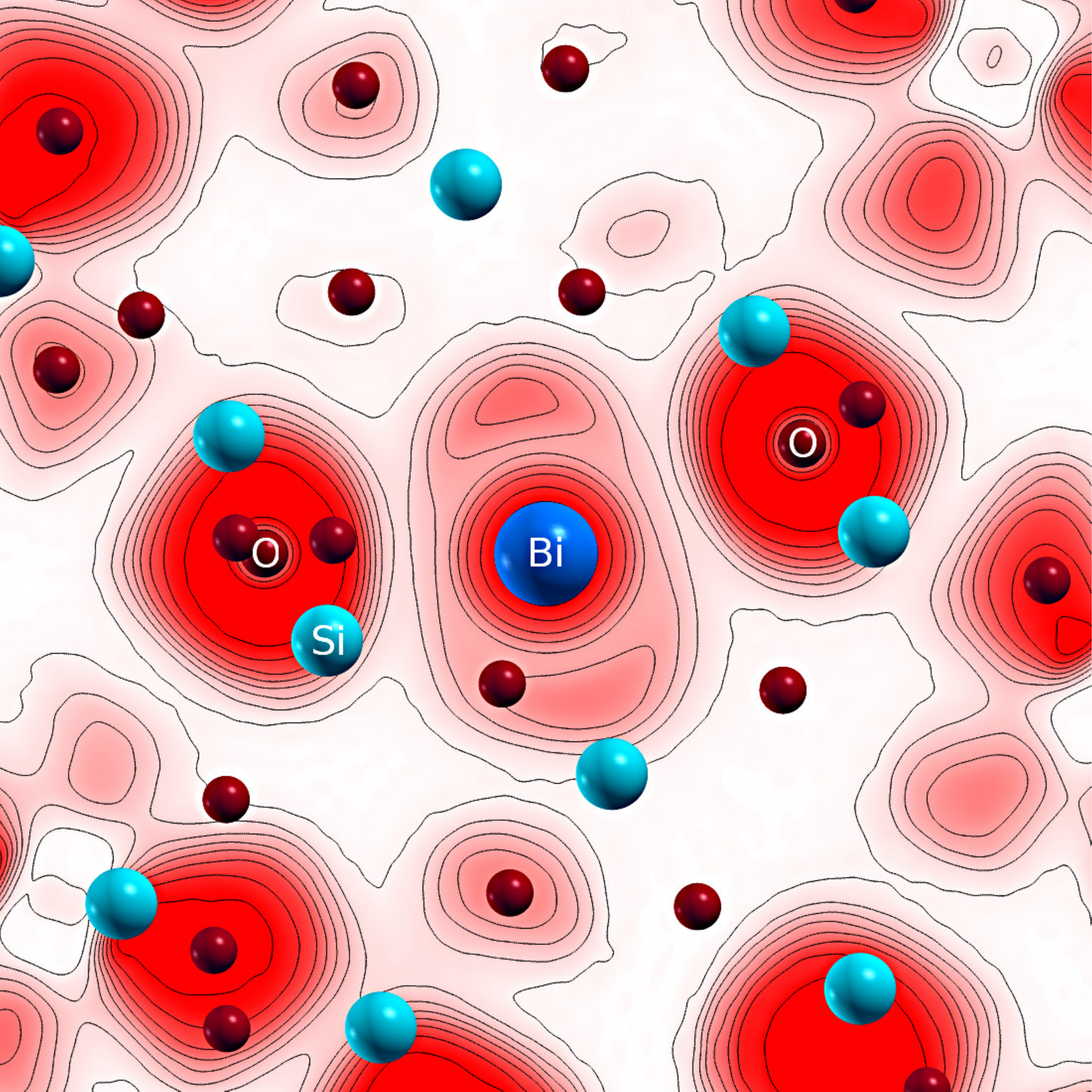}
\label{fig:Bi0_SiO2_ELF}
}
\subfigure[]{%
\includegraphics[width=8.75cm, bb=-10 -10 1110 1110]{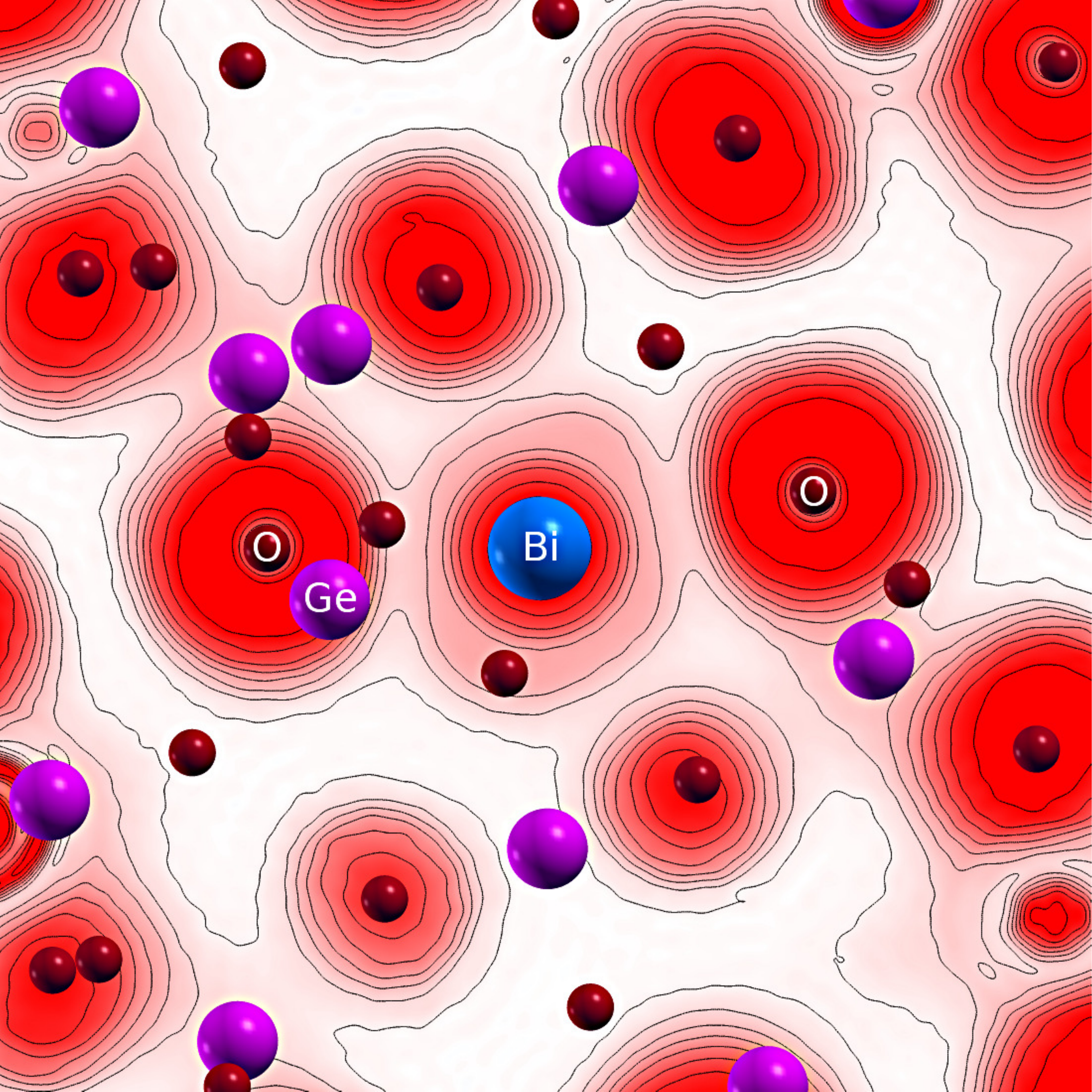}
\label{fig:Bi0_GeO2_ELF}
}
\\[-0.25\baselineskip]
\subfigure[]{%
\includegraphics[width=8.75cm, bb=-10 -10 1110 1110]{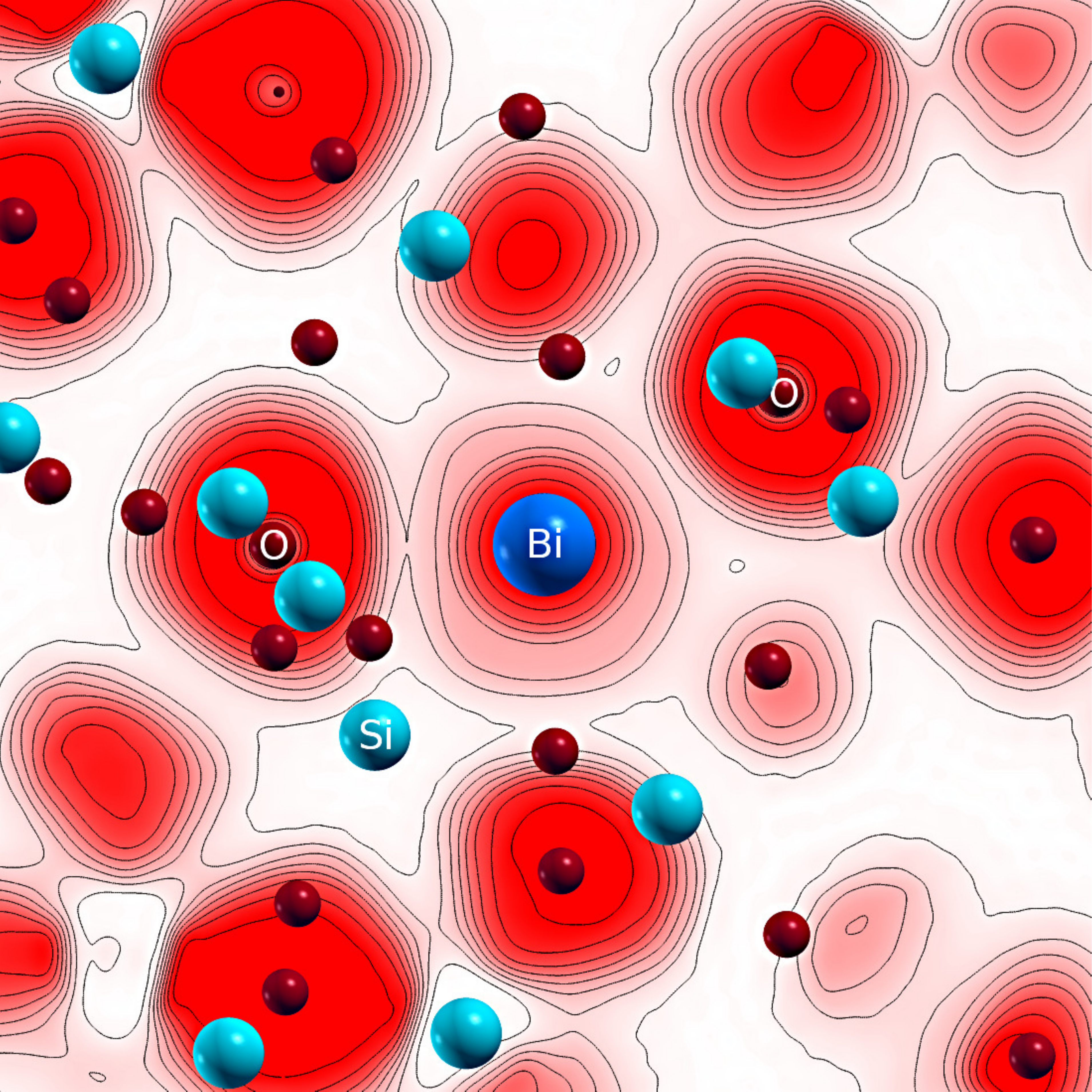}
\label{fig:Bi+_SiO2_ELF}
}
\subfigure[]{%
\includegraphics[width=8.75cm, bb=-10 -10 1110 1110]{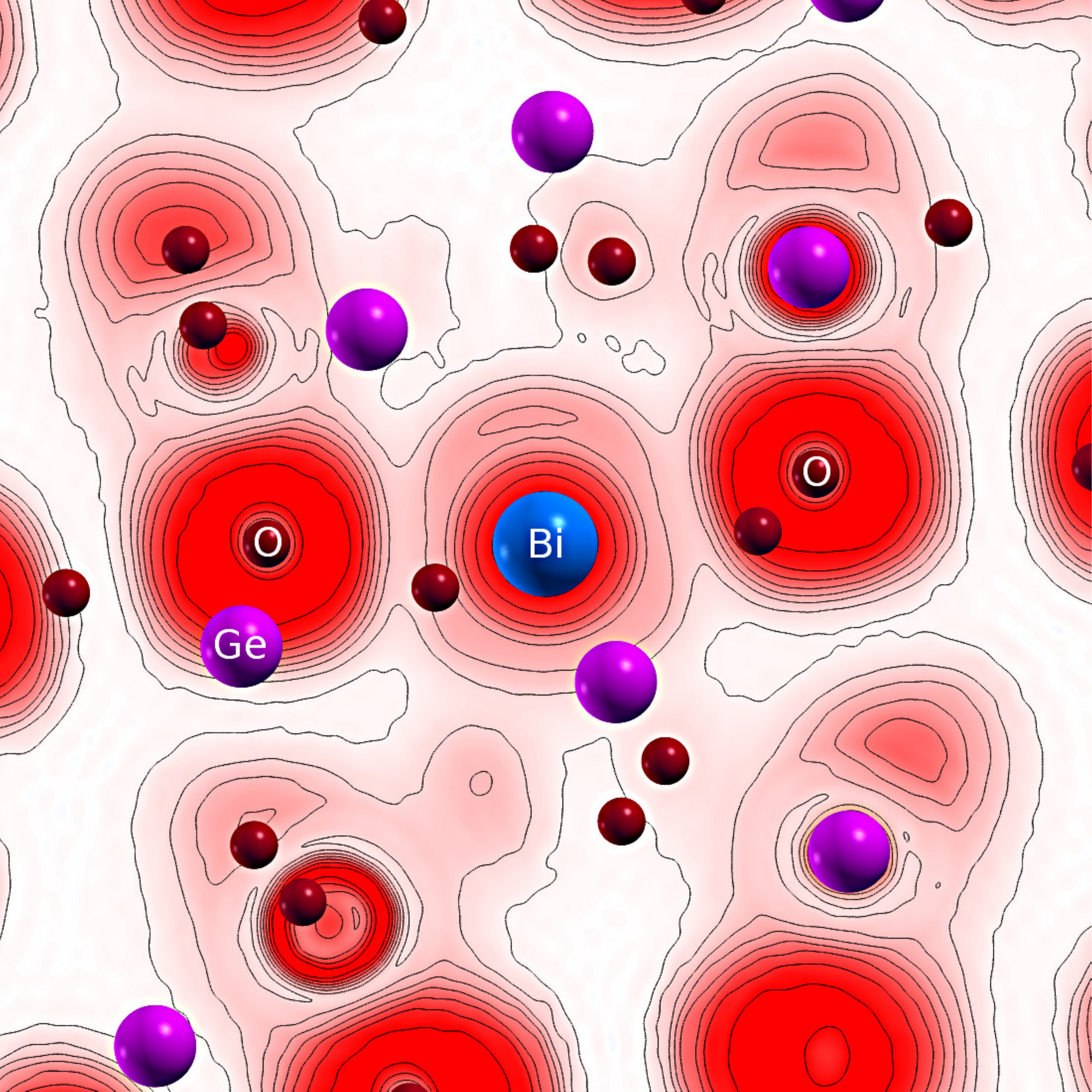}
\label{fig:Bi+_GeO2_ELF}
}
\vspace{1ex}
\caption{%
Calculated electron density maps of the centers formed by Bi interstitials:
\subref{fig:Bi0_SiO2_ELF}~\Biz{} atom in \SiOii,
\subref{fig:Bi0_GeO2_ELF}~\Biz{} atom in \GeOii,
\subref{fig:Bi+_SiO2_ELF}~\Bipi{} ion in \SiOii,
\subref{fig:Bi+_GeO2_ELF}~\Bipi{} ion in \GeOii.
The map plane goes through Bi atom and two nearest O atoms in each case.
}
\label{fig:Bi_ELF}
\end{figure*}

\subsection{%
Interstitial Bi$^0$ atom and Bi$^+$ ion
}
\label{sec:Bi^+_Bi^0}
According to the calculations, both \Biz{} atom and \Bipi{} ion can occur as
interstitial centers in six-member ring interstitial sites both in \SiOii{} and
\GeOii{} hosts (Fig.~\ref{fig:Bi_ELF}). The Bi interstitial atom or ion turn out
to interact weakly with the surrounding atoms and, as may be seen from
Fig.~\ref{fig:Bi_ELF}, do not form any bond with them. This means that the
influence of the host on electronic states and spectral properties of
interstitial Bi atom and ion consists mainly in crystal field effect.

So the spectral properties of the interstitials may be understood in a
model similar to the theory of \mbox{$\textrm{Tl}^0\!\left(1\right)$} center
\cite{Mollenauer83}. A weak axial crystal field is caused by O and Si ions of
SiO$_4$ six-member rings surrounding the interstitial center.
\begin{figure*}
\subfigure[]{%
\includegraphics[width=8.75cm, bb=-10 -10 1110 1110]{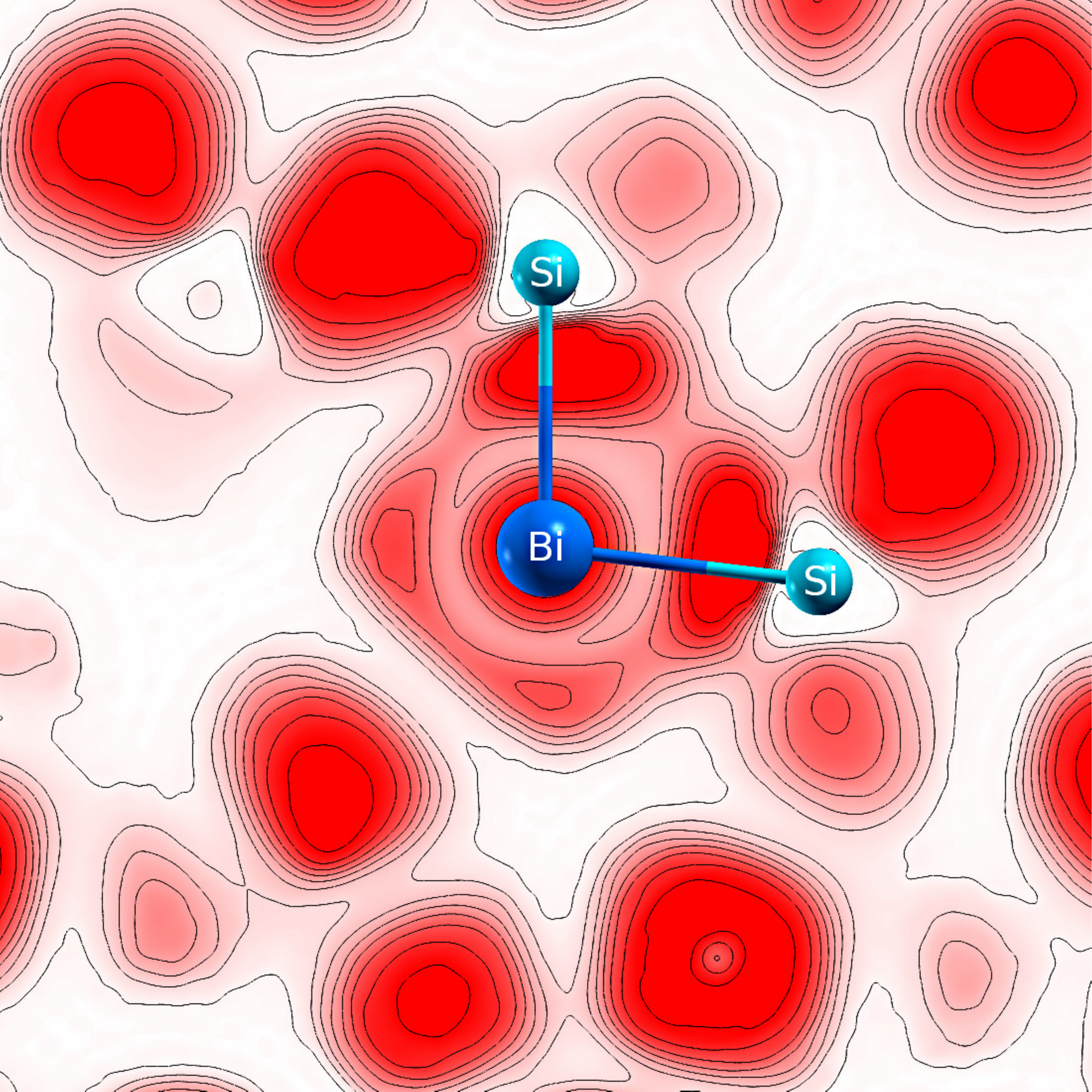}
\label{fig:Bi0_SiSi_ELF}
}
\subfigure[]{%
\includegraphics[width=8.75cm, bb=-10 -10 1110 1110]{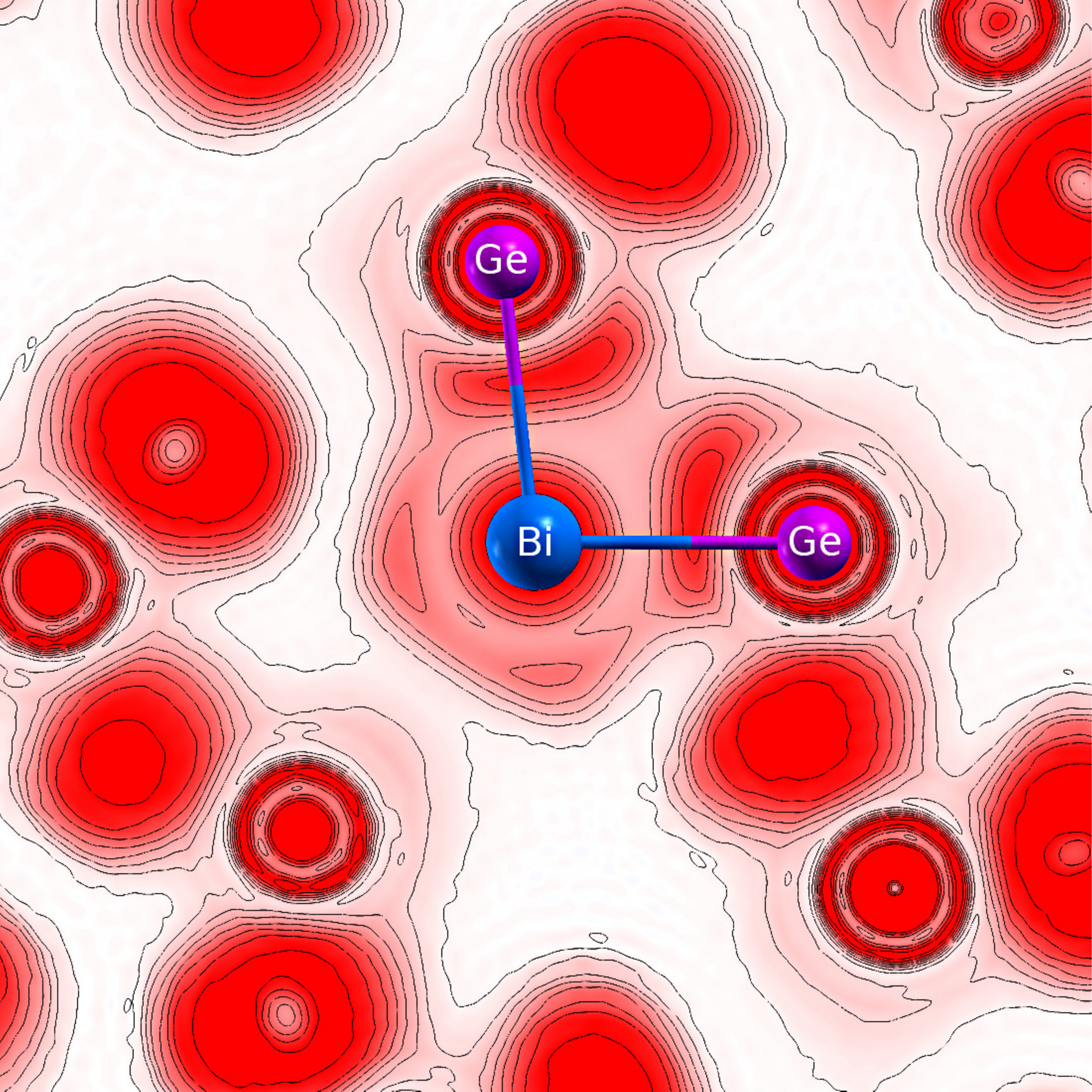}
\label{fig:Bi0_GeGe_ELF}
}
\caption{%
Calculated electron density maps of the centers formed by Bi interstitial atom
and oxygen vacancy: \subref{fig:Bi0_SiSi_ELF}~\Biz{} and \SiSi{} vacancy in
\SiOii, \subref{fig:Bi0_GeGe_ELF}~\Biz{} and \GeGe{} vacancy in \GeOii. The map
plane goes through the Bi atom and two Si or Ge atoms forming the vacancy in
each case.
}
\label{fig:Bi0_vacancy_ELF}
\end{figure*}

Three lowest states of \Bipi{} ion are known to arise from \Term{3}{P}{}{}
atomic state split by a strong spin-orbital interaction \cite{Moore58, George85,
Wahlgren01, Dolk02}. The ground state of \Bipi{} ion, \Term{3}{P}{0}{}, is not
split by the crystal field. The first excited state, \Term{3}{P}{1}{},
is split by an axial crystal field in two levels (approximately 10100 and
12300~\cminv{} in \SiOii{} and 10100 and 11400~\cminv{} in \GeOii), and the
second excited state, \Term{3}{P}{2}{}, is split in three levels (approximately
15500, 16600 and 20800~\cminv{} in \SiOii{} and 13500, 15400 and 18000~\cminv{}
in \GeOii), as displayed in Figs.~\ref{fig:Bi+_SiO2_levels} and
\ref{fig:Bi+_GeO2_levels}, respectively. Electric dipole transitions between
three spin-orbital components of \Term{3}{P}{}{} state forbidden in a free
\Bipi{} ion become allowed due to state mixing under the influence of the
crystal field. Levels and transitions schemes of \Bipi{} interstitial ions in
\SiOii{} and \GeOii{} are given in Figs.~\ref{fig:Bi+_SiO2_levels} and
\ref{fig:Bi+_GeO2_levels}, respectively.

The ground state of \Biz{} atom, \Term{4}{S}{3/2}{}, and the first,
\Term{2}{D}{3/2}{}, and second, \Term{2}{D}{5/2}{}, excited states are split by
the axial crystal field. The third excited state, \Term{2}{P}{1/2}{}, cannot be
split by the electrostatic field. Electric dipole transitions from the ground
state of \Biz{} interstitial to all the states arising from \Term{2}{D}{}{}
atomic state, turn out to be weak since in a free atom such
transitions are parity-forbidden. Hence the only relatively intensive absorption
band corresponding to  \Term{4}{S}{3/2}{}$\,\longrightarrow\,$\Term{2}{P}{1/2}{}
transition is expected to occur in \Biz{} center, in the $\lesssim 0.4$~\mkm{}
range, as follows from our calculations.

\subsection{%
Complexes formed by interstitial Bi atom and oxygen vacancy
}
\label{sec:Bi_and_O_vacancy}
According to our calculations, interstitial Bi atoms, \Biz, can form complexes
with intrinsic defects, \SiSi{} or \GeGe{} oxygen vacancies, in \SiOiiBi{} and
in \GeOiiBi, respectively.

We modeled oxygen vacancies, \SiSi{} in \SiOii{} and \GeGe{} in \GeOii, using
the same models and calculation approach as described above. In single \SiSi{}
vacancy the distance between Si atoms is found to be 0.244~nm, and in single
\GeGe{} vacancy the distance between Ge atoms is 0.258~nm. In both vacancies
a covalent bond between Si or Ge atoms is formed, somewhat relaxed in comparison
with crystalline Si or Ge. When a complex is formed with interstitial Bi atom,
the distance between Si atoms in \SiSi{} vacancy increases to 0.381~nm, and the
distance between Ge atoms in \GeGe{} vacancy increases to 0.391~nm. The
distances between the Bi atom and Si (Ge) atoms turn out to be 0.254 and
0.262~nm, respectively. Bader's analysis of electron density proves the
effective charge of Bi atom to be $+0.511\left|\textrm{e}\right|$ and
$+0.616\left|\textrm{e}\right|$ in \SiOiiBi{} and \GeOiiBi, respectively.
Effective charges of Si (Ge) atoms in the vacancy are found to be
$+3.020\left|\textrm{e}\right|$ and $+1.661\left|\textrm{e}\right|$,
respectively. In single \SiSi{} and \GeGe{} vacancies the effective charges of
Si and Ge atoms calculated using the same approach are
$+3.006\left|\textrm{e}\right|$ and $+1.863\left|\textrm{e}\right|$,
respectively. So an extra charge $\approx +0.54\left|\textrm{e}\right|$ or
$\approx +1.02\left|\textrm{e}\right|$ turns out to be localized in Bi atom and,
respectively, in Si or Ge atoms of \SiSi{} or \GeGe{} vacancy. Hence in such
complexes Bi turns out to occur in nearly monovalent state. With the complex
being formed, the electron density is redistributed. An effective charge
$\approx -0.46\left|\textrm{e}\right|$ and $\approx
-0.98\left|\textrm{e}\right|$ is transferred from Bi and Si (Ge) atoms into the
area between these Bi and Si (Ge) atoms and, to a lesser extent, into the area
between two Si (Ge) atoms. Corresponding electron density distributions are
shown in Figs.~\ref{fig:Bi0_SiSi_ELF} and \ref{fig:Bi0_GeGe_ELF} in
\mbox{$\textrm{Si}\!\relbar\!\textrm{Bi}\!\relbar\!\textrm{Si}$} or
\mbox{$\textrm{Ge}\!\relbar\!\textrm{Bi}\!\relbar\!\textrm{Ge}$} planes. Thus,
three-center mutually bound group of Bi atom and two Si (Ge) atoms is formed
instead of a pair of covalently bonded Si (Ge) atoms. Coordination-type
three-center bond would be expected to occur in such a complex. It can be seen
in Figs.~\ref{fig:Bi0_SiSi_ELF} and \ref{fig:Bi0_GeGe_ELF} that Bi atoms remains
substantially isolated and do not form pronounced two-center bonds with Si or Ge
atoms.

Levels and transitions schemes of \BiVSi{} and \BiVGe{} complexes in \SiOiiBi{}
and \GeOiiBi{} are shown in Figs.~\ref{fig:Bi0_SiSi_levels} and
\ref{fig:Bi0_GeGe_levels}, respectively.

Again, the spectral properties of such a complex may be understood in a
crystal-field model. Basing on the described rearrangement of the electron
density, one may consider the complex in a rough approximation as a pair of
centers, an interstitial \Bipi{} ion and a negatively charged
\mbox{$\equiv\!\textrm{Si}
{\overset{{}^\bullet}{\relbar}} \textrm{Si}\!\equiv$}
(\mbox{$\equiv\!\textrm{Ge}
{\overset{{}^\bullet}{\relbar}}\textrm{Ge}\!\equiv$}) vacancy, similar
to thallium and lead centers in crystalline alkali and alkali-earth halides
(see e.g. discussion in \cite{Dianov10}) or to a complex formed by bismuth
substitutional center and anion vacancy in TlCl crystal \cite{Plotnichenko13,
Sokolov13}. In other words, the center is considered to be a \Bipi{} ion in
axial crystal field formed by a neighboring negatively changed oxygen vacancy.
The ground state of \Bipi{} ion, \Term{3}{P}{0}{}, is not split. The first
excited state, \Term{3}{P}{1}{}, is split by axial crystal field in two levels,
approximately 7000 and 7900~\cminv{} in \SiOii{} and 5900 and 6600~\cminv{} in
\GeOii. The second excited state, \Term{3}{P}{2}{}, is split in three levels,
approximately 11000, 12200, and 24400~\cminv{} in \SiOii{} and 9500, 10500, and
22600~\cminv{} in \GeOii. The next excited state of \Bipi{} ion,
\Term{1}{D}{2}{}, is split in three levels as well, energy of the lowest of
those being about 27000--28000~\cminv{} and other two lie at
41000--45000~\cminv. One more excited state of \Bipi{} ion, \Term{1}{S}{0}{},
not split by the crystal field, occurs somewhat above 45000~\cminv{} in \SiOii{}
or 46000~\cminv{} in \GeOii{} (Fig.~\ref{fig:Bi_centers_levels},
\subref{fig:Bi0_SiSi_levels} and \subref{fig:Bi0_GeGe_levels}). Electric dipole
transitions between the ground state and the split excited ones become allowed
due to state mixing in the crystal field.

It should be realized that whilst such a model is useful to understand
qualitatively the origin of IR luminescence in \BiVSi{} and \BiVGe{} complexes,
it nevertheless provides only very approximate description of their
electronic structure. In fact, changes in electronic states and spectral
properties of Bi ion in \BiVSi{} and \BiVGe{} complexes go far beyond the
crystal field effect in comparison with both free \Bipi{} ion and the
above-considered interstitial \Bipi{} ion. The coordination-bond-type
distribution of electron density covers both the Bi atom and two Si (Ge) atoms
in the oxygen vacancy. This results in Bi ion states split and mixed
considerably stronger than it is even possible in the crystal field (e.g., in
the case of interstitial \Bipi{} ions). As a result, the corresponding
transition wavelengths turn out to be increased considerably. In \GeOii{} the
electron density redistribution is more pronounced, and transitions in \BiVGe{}
center turn out to be shifted to larger wavelengths than in \BiVSi{} one. On
the other hand, since no distinct diatomic covalent bonds,
\mbox{$\textrm{Bi}\!\relbar\!\textrm{Si}$} or
\mbox{$\textrm{Bi}\!\relbar\!\textrm{Ge}$}, are formed in the complexes, the
electronic states of \BiVSi{} and \BiVGe{} centers turn out to depend very
slightly on small displacements of Bi atom. Hence Stokes shift is expected to be
low, at least for the most long-wave transitions.
\begin{figure}
\includegraphics[width=8.75cm, bb=5 0 405 405]{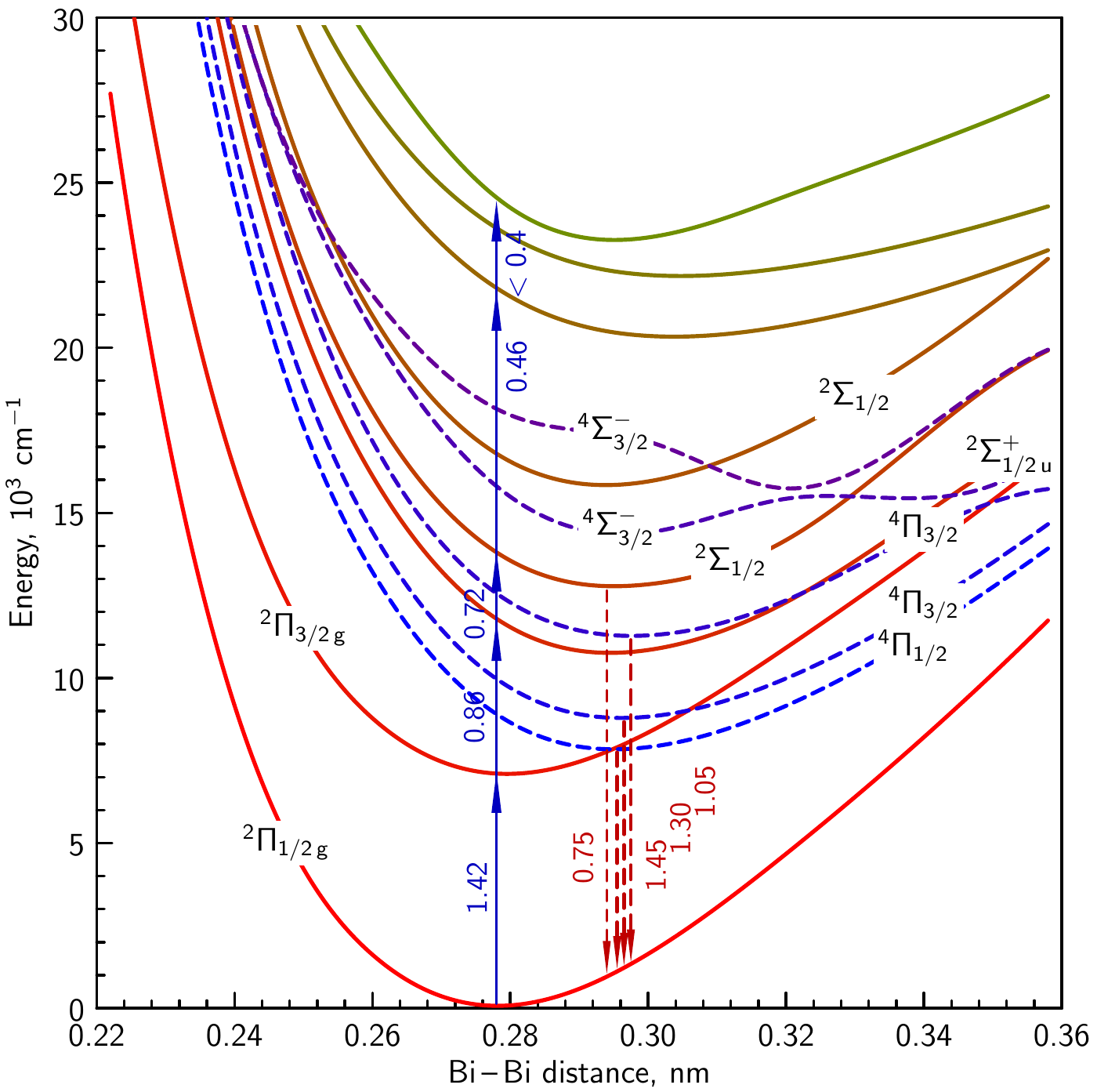}
\caption{%
\label{fig:Bi_2^-_levels}
Total energy curves and transitions in \Biiim{} dimer
}
\end{figure}

\subsection{%
Interstitial Bi dimers
}
\label{sec:Bi_2^-}
The modeling proves both neutral, \Biiiz, and negative single-charged, \Biiim,
bismuth dimers to occur in \SiOii{} network in stable positions in interstitial
sites formed by six-member rings of SiO$_4$ tetrahedra. Similar to
\BiO{} molecule discussed above, \Biiiz{} or \Biiim{} dimer lie in such
equilibrium positions between two adjacent rings being aligned along the ring
interstitial axis. Again, either forced declination from the axis or shift aside
do not lead to any reaction of the dimer with the surrounding atoms. Recently
such stable positions of Bi dimers were found in aluminosilicate glasses in our
calculations performed by quantum-chemical methods in cluster models of the
networks \cite{Sokolov09}. On the contrary, only neutral interstitial dimer,
\Biiiz, is found to be stable in \GeOii{} network. The negatively charged dimer,
\Biiim, turns out to be unstable, and once placed into the ring interstitial
site, readily enters into reaction with the neighboring atoms. As a result,
three- and fourfold coordinated Bi atoms (in other words, trivalent Bi
substitutional centers discussed in Section~\ref{sec:Bi^3+}) are formed.
\begin{figure*}
\subfigure[]{%
\includegraphics[width=8.75cm, bb=-20 -10 310 800]{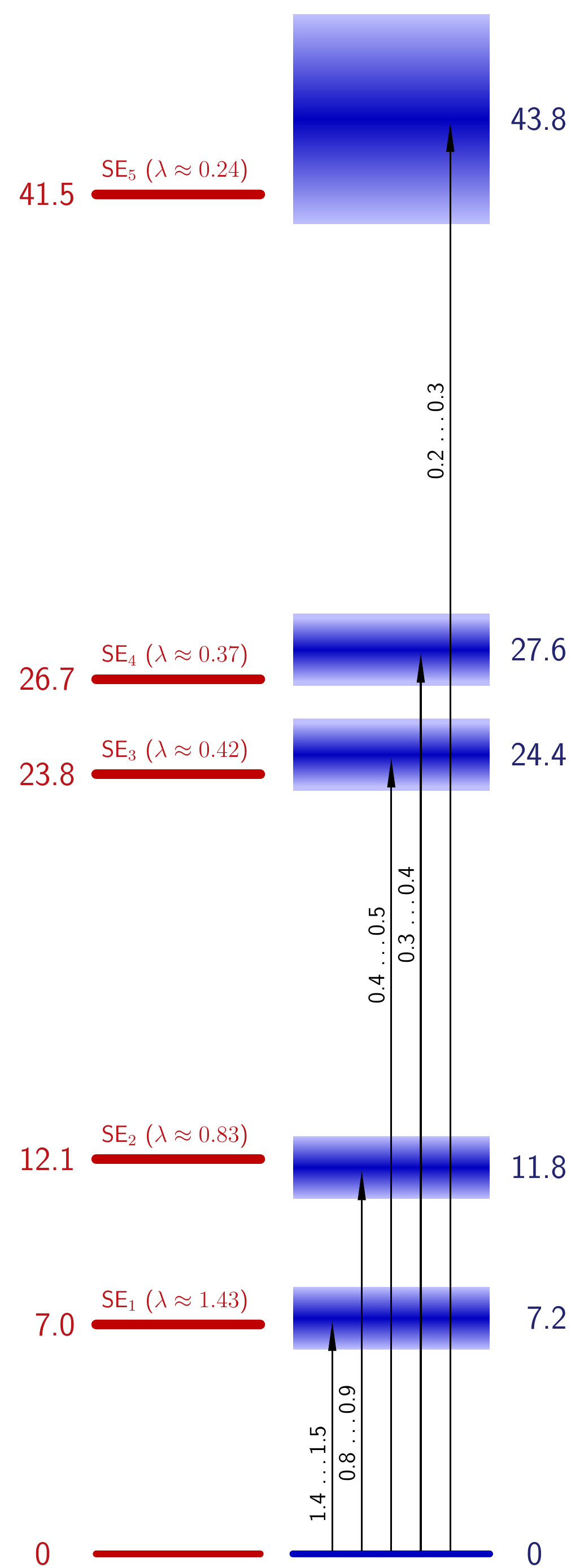}
\label{fig:Bi0_SiSi_vs_Bufetov}
}
\subfigure[]{%
\includegraphics[width=8.75cm, bb=-20 -10 310 800]{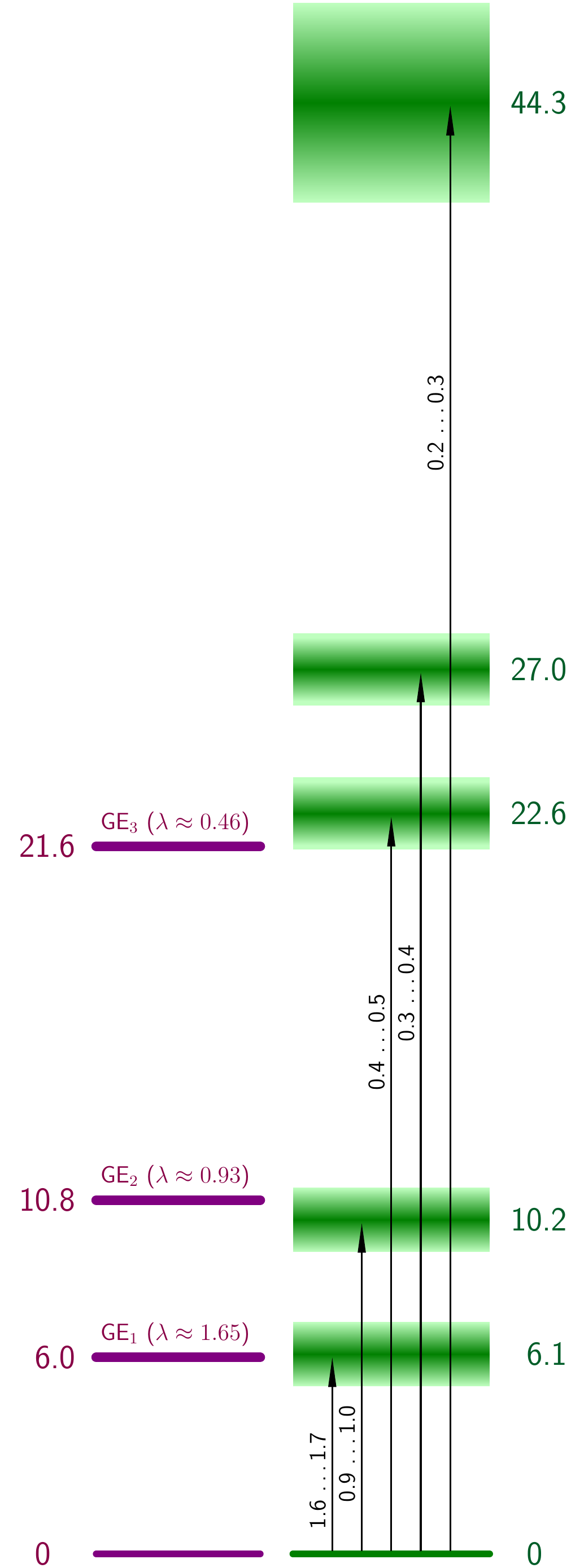}
\label{fig:Bi0_GeGe_vs_Bufetov}
}
\caption{%
Comparison of the calculated levels and transitions schemes of centers formed by
Bi interstitial atom and oxygen vacancy with the empirical schemes of
bismuth-related centers in \SiOii{} and \GeOii{} suggested in
Refs.~\cite{Bufetov11b, Bufetov12a, Bufetov12b}:
\subref{fig:Bi0_SiSi_vs_Bufetov}~\Biz{} and \SiSi{} vacancy in \SiOii{} vs.
\mbox{Si-BAC} center;
\subref{fig:Bi0_GeGe_vs_Bufetov}~\Biz{} and \GeGe{} vacancy in \GeOii{} vs.
\mbox{Ge-BAC} center.
Level energies are given in $10^3$~\cminv, transition wavelengths in \mkm.
}
\label{fig:Bi0_vacancy_vs_Bufetov}
\end{figure*}
\begin{table*}
\caption{%
Estimated relative lifetimes of luminescence in bismuth-related centers
}
\begin{center}
\begin{tabular}{cccccccc}
\hline\hline
&&&&&&                                           \\[-1.25ex]
luminescence & \Bipiii & \Bipii & \BiO & \Bipi &
\multicolumn{2}{c}{Bi $\cdots$ vacancy} & \Biiim \\
&&&&& \SiSi & \GeGe &                            \\[1.25ex]
\hline
&&&&&&& \\[-1.25ex]
near-IR      & --- & ---                     & $\gtrsim 10^2$             &
$\sim 10^2$                & $\gtrsim 10^2$             &
$\sim 10^2$                & $\gtrsim 10^2$                           \\[1.00ex]
visible / IR & 1   & $\sim 10^{\phantom{2}}$ & $\gtrsim 10^{\phantom{2}}$ &
$\gtrsim 10^{\phantom{2}}$ & $\gtrsim 10^{\phantom{2}}$ &
$\sim 10^{\phantom{2}}$    & $\sim 10^{\phantom{2}}$                  \\[1.50ex]
\hline\hline
\end{tabular}
\end{center}
\label{tab:lifetimes}
\end{table*}

Spectral properties of \Biiim{} dimer have been studied in enough detail, for
the first time in Ref.~\cite{Balasubramanian91} and later in
Refs.~\cite{Sokolov09, Sokolov08}, by configuration interaction method in
various approximations. The calculated total energy curves for several low-lying
states of \Biiim{} dimer are shown in Fig.~\ref{fig:Bi_2^-_levels} (see
as well Fig.~3 in \cite{Balasubramanian91}, Fig.~1 in \cite{Sokolov09}, and
Fig.~1 in \cite{Sokolov08}). Transitions and corresponding absorption and
possible luminescence in \Biiim{} dimers are discussed in detail in
Refs.~\cite{Sokolov08, Sokolov09}.

\section{%
Discussion
}
\label{sec:Discussion}
Comparing the experimental data reported in Refs.~\cite{Bufetov11a, Bufetov11b,
Bufetov12a, Bufetov12b, Bufetov13, Trukhin13b} (see
Section~\ref{sec:Introduction}) with the above-described results of our
calculations, one may safely suggest that the bismuth-related centers giving
rise to IR luminescence in \SiOiiBi{} and \GeOiiBi{} glasses are likely to be,
respectively, \BiVSi{} and \BiVGe{} complexes formed by interstitial Bi atoms
and oxygen vacancies. Indeed, the calculated spectral properties of these
complexes are in good agreement with the experimental data \ref{itm:i}i --
\ref{itm:v} (page~\pageref{itm:v}). According to the calculations, in \SiOiiBi{}
bismuth-related absorption is expected in two broad bands near 1.4 and 0.8~\mkm,
and in two bands near 0.4 and 0.35~\mkm. Broad-band near-IR luminescence near
1.4~\mkm{} is expected to be excited in all these absorption bands, and the IR
luminescence in the 0.8--0.9~\mkm{} range is expected to be excited in the
latter three absorption bands. In \GeOiiBi{} bismuth-related absorption is
expected in broad bands near 1.6 and 1.0~\mkm, and in the bands near 0.45 and
0.3~\mkm. Near-IR luminescence in a broad band near 1.6~\mkm{} is expected
to be excited in all these absorption bands, and the IR luminescence in the
0.9--1.0~\mkm{} range is expected to be excited in the latter three absorption
bands. As evident from all these data displayed in
Fig.~\ref{fig:Bi0_vacancy_vs_Bufetov}, the calculated levels and transitions
schemes of \BiVSi{} and \BiVGe{} centers accord closely with the empirical
schemes proposed in Refs.~\cite{Bufetov11b, Bufetov12a, Bufetov12b}. As well
the suggested model of bismuth-related centers provides a support for the
above-mentioned hypothesis concerning a common origin and similar structure of
near-IR luminescence centers in bismuth-doped \SiOii{} and \GeOii{} glasses
\cite{Bufetov10, Bufetov11a, Bufetov11b, Bufetov12a, Bufetov12b, Bufetov13}.

On the other hand, interstitial \BiO{} molecules and interstitial negatively
changed \Biiim{} dimers could contribute to IR luminescence in \SiOiiBi{} as
well. Spectral properties of \BiO{} molecule were discussed in detail in
Ref.~\cite{Sokolov11} specifically in relation to interstitial \BiO{} molecule
in \SiOii{} glass, and were shown to agree well with the experimental absorption
and luminescence spectra observed in \SiOiiBi{} \cite{Bufetov10, Bufetov11a,
Bufetov11b}. In general, spectral properties of \Biiim{} dimer agree
satisfactorily with these experimental spectra \cite{Sokolov09, Sokolov08}, with
two additional remarks: the luminescence corresponding to the transition near
1.300~\mkm{} might not be observed in experiment due to low excitation
efficiency, and the absorption and luminescence in the 0.82--0.83~\mkm{} range
may be ascribed to the transitions near 0.72~\mkm{} and near 0.86~\mkm,
respectively (see Fig.~\ref{fig:Bi_2^-_levels} and Figures~1 in
Refs.~\cite{Sokolov09, Sokolov08}). However, according to our modeling, neither
interstitial \BiO{} molecules nor interstitial negatively changed \Biiim{}
dimers can occur in \GeOiiBi. Thus, among all the bismuth-related centers
studied, \BiVSi{} and \BiVGe{} complexes seem to be the most appropriate model
of IR luminescence bismuth-related centers of the same origin both in \SiOii{}
and \GeOii.

In general, the results of our modeling support the suggestions concerning the
origin of red luminescence in \SiOiiBi{} glasses made in Refs.~\cite{Bufetov11a,
Bufetov11b, Trukhin13b}, and allow one to understand the absence of such
luminescence in \GeOiiBi{} glasses. However it seems reasonable to discuss
certain suggestions \cite{Trukhin13b} in more detail.

According to Ref.~\cite{Trukhin13b}, there are luminescence centers of two types
in \SiOiiBi. Bi centers of the first type are responsible for IR luminescence
near 1.4~\mkm{} with a long lifetime excited only by intra-center absorption.
Centers of the second type give rise to red luminescence near 0.63~\mkm{} with a
short lifetime excited not only by intra-center absorption but by electron-hole
recombination as well. Such centers trap electrons excited to the conduction
band by two-photon absorption of intense UV laser light, and the recombination
of trapped electrons with holes gives rise to red luminescence. The centers
responsible for IR luminescence are not quenched up to the temperature of 700~K,
but the red luminescence centers are thermally quenched at temperatures $\gtrsim
450$~K (\ref{itm:vii} in page~\pageref{itm:vii}). Such different properties of
bismuth-related centers might be explained by the fact that the first-type
center weakly interacts with the glass network and, as a result, it is
temperature-resistant and cannot capture conductive electrons and participate in
the process of electron-hole recombination excitation of the luminescence. The
second-type center is embedded in the glass network and hence it is subject to
temperature quenching and is able to capture conductive electrons with
subsequent excitation of red luminescence owing to electron-hole recombination.

The results of modeling bismuth-related centers described in
Sections~\ref{sec:Bi^2+} and \ref{sec:Bi_and_O_vacancy} allow an understanding
the assumptions made by the authors of Ref.~\cite{Trukhin13b} on the basis
of their experimental data.

Indeed, if the IR luminescence center in \SiOiiBi{} is a complex of interstitial
Bi atom and \SiOii{} intrinsic defect, \SiSi{} oxygen vacancy, its thermal
stability should be only somewhat lower than the stability of the vacancy, since
in such a complex the Bi atom forms only weak bonds with the glass network.
Further, these bonds are formed with Si atoms of the vacancy, but not with O
atoms. This eliminates the possibility of recombination of an electron trapped
in such a center with a hole in the glass valence band, since in
\SiOii{} the states near the upper edge of the valence band are formed mainly by
non-bonding 2p electron states (lone pairs) of bridging O atoms. On the other
hand, if the red luminescence center in \SiOiiBi{} is the divalent Bi
substitutional atom, \Bipii, with the Bi atom covalently bonded with two
bridging O atoms, the thermal stability of such a center should be considerably
lower than the stability of the vacancy owing to low coordination of the Bi
atom. The bridging O atoms in this center allow effective recombination of a
hole with a trapped electron.

To realize the possibility of conduction electron to be captured in the defects
under consideration we calculated electron affinities of \BiVSi{} complex and of
\Bipii{} substitutional center using the approach developed in
Ref.~\cite{Foster02}. The electron affinity is found to be approximately of 2.1
and 4.2~eV for these centers, respectively. According to different estimations
\cite{Fischetti87, Trukhin89, Wang96}, electron affinity of vitreous \SiOii{}
host is known to be from 0.9 to 2.7~eV. Hence our calculations suggest that
conduction electron can be readily trapped in divalent Bi substitutional center,
\Bipii, and can hardly be trapped in \BiVSi{} center. This is in obvious
agreement with the assumptions made in Ref.~\cite{Trukhin13b}.

As noted above, comparing the results of lifetime calculations with the
experimental data on the excited states responsible for the luminescence in
\SiOiiBi{} and \GeOiiBi{} (\ref{itm:viii} in page~\pageref{itm:viii}) is
meaningful only by an order of magnitude. For this purpose the calculated
relative lifetimes are listed in Table~\ref{tab:lifetimes} with the lifetime
corresponding to the most intensive transition in \Bipiii{} substitutional
center (the absorption near 0.25~\mkm, the luminescence in the 0.3--0.4~\mkm{}
range, see Section~\ref{sec:Bi^3+}) taken to be unity. The lifetime of
corresponding luminescence of trivalent Bi in different hosts is known to be
1--5~\mks{} \cite{Blasse68, Blasse84}. Using this estimation together with the
data given in Table~\ref{tab:lifetimes} one can easily see that the results of
lifetime calculations agree satisfactorily with the experimental data both for
IR luminescence of \BiVSi{} and \BiVGe{} centers and red luminescence of
\Bipii{} substitutional center.

\section{%
Conclusion
}
\label{sec:Conclusion}
Basing on the results of our modeling of bismuth-related centers in \SiOiiBi{}
and \GeOiiBi{} glasses and taking into account the assumption concerning
common origin of Bi centers responsible for IR luminescence in silica and
germania glasses, it may be safely suggested that the bismuth-related centers of
near-IR luminescence in \SiOiiBi{} and \GeOiiBi{} glasses are mainly represented
by the \BiVSi{} and \BiVGe{} complexes formed by interstitial Bi atoms and
intrinsic defects of glass, \SiSi{} and \GeGe{} oxygen vacancies. Interstitial
\BiO{} molecules and negatively charged \Biiim{} dimers might also contribute to
the IR luminescence in \SiOiiBi{} glass as well but are absent in \GeOii{} host.
Bismuth-related centers responsible for the visible (red) luminescence in the
\SiOiiBi{} glass are likely to be represented by twofold coordinated Bi atoms
bonded with Si atoms by bridging O atoms.

\section*{%
Acknowledgments
}
The authors are grateful to Prof.~I.A.Bufetov for valuable discussions. This
work is supported in part by Basic Research Program of the Presidium of the
Russian Academy of Sciences.
%
%

\end{document}